\documentclass{elsart}

\usepackage{graphicx}

\newcommand{\lepton}{\ifmmode {l} \else $l$\fi}
\newcommand{\elepton}{\ifmmode {l^{*}} \else $l^{*}$\fi}
\newcommand{\wboson}{\ifmmode {{\mathrm W}^{\pm}} \else
${\mathrm W}^{\pm}$\fi}
\newcommand{\wpair}{\ifmmode {{\mathrm W}^{+}{\mathrm W}^{-} } \else
${\mathrm W}^{+}{\mathrm W}^{-}$\fi}
\newcommand{\zboson}{\ifmmode {{\mathrm Z}^{0}} \else
${\mathrm Z}^{0}$\fi}

\usepackage{amssymb}
\usepackage{amsmath}
\usepackage{mathrsfs}
\usepackage{epsfig}

\begin{document}

\begin{frontmatter}



\title{The sensitivity of cosmic ray air shower experiments for excited lepton detection}


\author[lip]{M.C. Esp\'{\i}rito Santo}
\author[lip]{M. Paulos}
\author[lip,ist]{M. Pimenta \corauthref{cor}}
\author[ist]{J. C. Rom\~ ao}
\author[lip]{B. Tom\'e\thanksref{grant}}
\address[lip]{LIP, Av. Elias Garcia, 14--1, 1000-149 Lisboa Portugal}
\address[ist]{IST, Av. Rovisco Pais, 1049-001 Lisboa, Portugal}
\corauth[cor]{pimenta@lip.pt, LIP, Av. Elias Garcia, 14--1, 1000-149 Lisboa Portugal.}
\thanks[grant]{FCT grant SFRH/BPD/11547/2002.}

\begin{abstract}
In models with substructure in the fermionic sector, excited fermion states are expected.
Excited leptons could be produced in the interaction of high energy quasi-horizontal 
cosmic neutrinos with the atmosphere via neutral and charged current processes,
$\nu N \rightarrow \nu^* X$ and $\nu N \rightarrow \ell^* X$.
The hadronic component X, and possibly part of the excited lepton decay products,
would originate an extensive air shower, observable in large cosmic
ray experiments. In this paper, the sensitivity of present and planned very high
energy cosmic ray experiments to excited lepton production is estimated and discussed.

\end{abstract}

\end{frontmatter}

\section{Introduction}
\label{sec:introd}

Compositeness is a never discarded hypothesis for explaining the complexity 
of the present fundamental particle picture.
In models with substructure in the fermionic sector, excited fermion states
are expected.
In the past years many searches were performed in the accelerators around the 
world~\cite{accel}. So far, no evidence for excited fermions was found, 
and stringent limits were set at the electroweak scale.

The race for higher energies has new partners in present and future large 
cosmic rays experiments.
These experiments, covering huge detection areas, are able to explore the 
high energy tail of the cosmic ray spectrum, reaching centre-of-mass 
energies orders of magnitude above those of man made accelerators. 
Although having poorer detection capabilities and large uncertainties on the beam 
composition and fluxes, cosmic ray experiments present a unique opportunity to 
look for new physics at scales far beyond the TeV. 

Energetic cosmic particles interact with the atmosphere of Earth originating 
Extensive Air Showers (EAS) containing billions of particles. While cosmic particles 
with strong or electromagnetic charges are absorbed in the first layers of the 
atmosphere, neutrinos have a much lower interaction cross-section and can easily 
travel large distances. 
Energetic cosmic neutrinos, although not yet observed and with very large 
uncertainties on the expected fluxes, are predicted on rather solid grounds~\cite{cosmneut}.
Nearly horizontal neutrinos, seeing a large target volume 
and with negligible background from ``ordinary'' cosmic rays, are thus an ideal beam 
to explore possible rare processes~\cite{exoticnu}.

In this paper, the possibility of excited lepton searches in current (AGASA~\cite{agasa}, 
Fly's Eye~\cite{fly}) and 
future (Auger~\cite{auger}, EUSO~\cite{euso}, OWL~\cite{owl}) very high energy cosmic ray 
experiments is discussed.
Excited leptons could be produced in the interaction of high energy quasi-horizontal
cosmic neutrinos with the atmosphere via neutral and charged current processes, 
$\nu N \rightarrow \nu^* X$ and $\nu N \rightarrow \ell^* X$
($\nu^*$ and  $\ell^*$ representing neutral and charged excited leptons, respectively). 
The hadronic component $X$, and possibly part of the excited lepton decay products, 
would originate an extensive air shower, observable by large cosmic ray experiments.
As the initial beam must contain all three neutrino flavours,
one expects the production of excited leptons of first, second and
third family.
In the specific case of the third family excited leptons decaying into $\tau X$, 
the subsequent decay of the tau lepton may originate a second visible air shower 
within the acceptance of the experiment and thus giving rise to a double bang event 
topology~\cite{dbang,bh-our}.
Excited leptons are hypothetical particles, and their unknown interactions are often
described at the electroweak scale by effective models. In this work the well known model 
of~\cite{hagiboud} is assumed. This model is described in section~\ref{sec:excited}, where the
excited lepton production cross-sections are obtained and the decay branching ratios at the 
relevant energies and masses are discussed.
In section~\ref{sec:sensitivities} the expected sensitivity to excited lepton events of 
the largest available and planned cosmic ray experiments is estimated and discussed.
Some conclusions are finally drawn.

\section{Production and decay of excited leptons}
\label{sec:excited}

The SU(2)$\times$U(1) gauge invariant effective  Lagrangian describing the  
magnetic transition between excited leptons and the Standard Model (SM) leptons has 
the form~\cite{hagiboud}: 
\begin{equation}
 {\mathscr L}_{\lepton \elepton} =                    
                    \frac{1}{2\Lambda} \overline{{L}^*} 
                    \sigma^{\mu \nu} 
                   \left[ 
                         g f \frac{{\boldsymbol \tau}}{2}  {\mathbf W}_{\mu \nu} +
                         g' f' \frac{Y}{2} B_{\mu \nu} 
                   \right] {L}_L       
                   + \text{h.c.}
\end{equation}
\noindent
where ${L}^*={L}^*_L + {L}^*_R $, with:

$$
{L}^*_L = \begin{bmatrix} \nu^* \\ \ell^*  \end{bmatrix}_L ; \quad 
{L}^*_R = \begin{bmatrix} \nu^* \\ \ell^*  \end{bmatrix}_R  
$$

\noindent
and $L_L$ is the weak isodoublet with the left-handed components of the SM leptons.
Above, $\sigma^{\mu \nu}$ is the covariant bilinear tensor, 
$\boldsymbol \tau$ are the Pauli matrices,  $Y$ is the weak hypercharge,  
${\bf W}_{\mu \nu}$ and $B_{\mu \nu}$ represent the gauge field tensors of 
SU(2) and U(1), respectively, and $g$ and $g'$ are the corresponding 
SM coupling constants.
The parameter $\Lambda$ sets the compositeness scale and 
$f$, $f'$ are weight factors associated with the two gauge groups.

This Lagrangian describes the $\lepton \elepton V$ vertex, and thus
the single production of excited leptons and their decays. 
The strength of the $\lepton \elepton V$ coupling is parameterised through 
$f$ and $f'$.
Form factors and anomalous magnetic moments of the excited leptons were not
considered. From this Lagrangian we can derive the vertex
$$
\Gamma_\mu^{Vl^*l}= \frac{e}{2\Lambda} q^\nu \sigma_{\mu\nu} (1-\gamma_5) f_V
$$
where the couplings to the physical gauge bosons are
$$
f_\gamma=e_l f' + I_{3L}(f-f'),~~~ f_W=\frac{1}{\sqrt 2 s_W}f, ~~~
f_Z=\frac{I_{3L} (c^2_W f + s^2_W f') -e_l s^2_W f'}{s_W c_W} 
$$
$I_{3L}$ being the fermion weak isospin and 
$s_W=\sin \theta_W, c_W=\cos\theta_W$, 
with $\theta_W$ the SM weak mixing angle.
To reduce the number of free parameters, it is customary to assume a relation between $f$
and $f'$. In this paper, the scenarios $f=f'$ and
$f=-f'$ will be considered. 
In these cases, the single excited lepton production cross-section depends only on the ratio 
$|f|/\Lambda$ and on the excited lepton mass. 
It is worth noting that
for $f=f'$ the coupling of the excited neutrinos to the photon vanishes. The same
is true for excited charged leptons if $f=-f'$.

\subsection{Production}
\label{sec:prod}

\begin{figure}[p]
\begin{center}
\vspace{-0.5cm}
\mbox{\epsfig{file=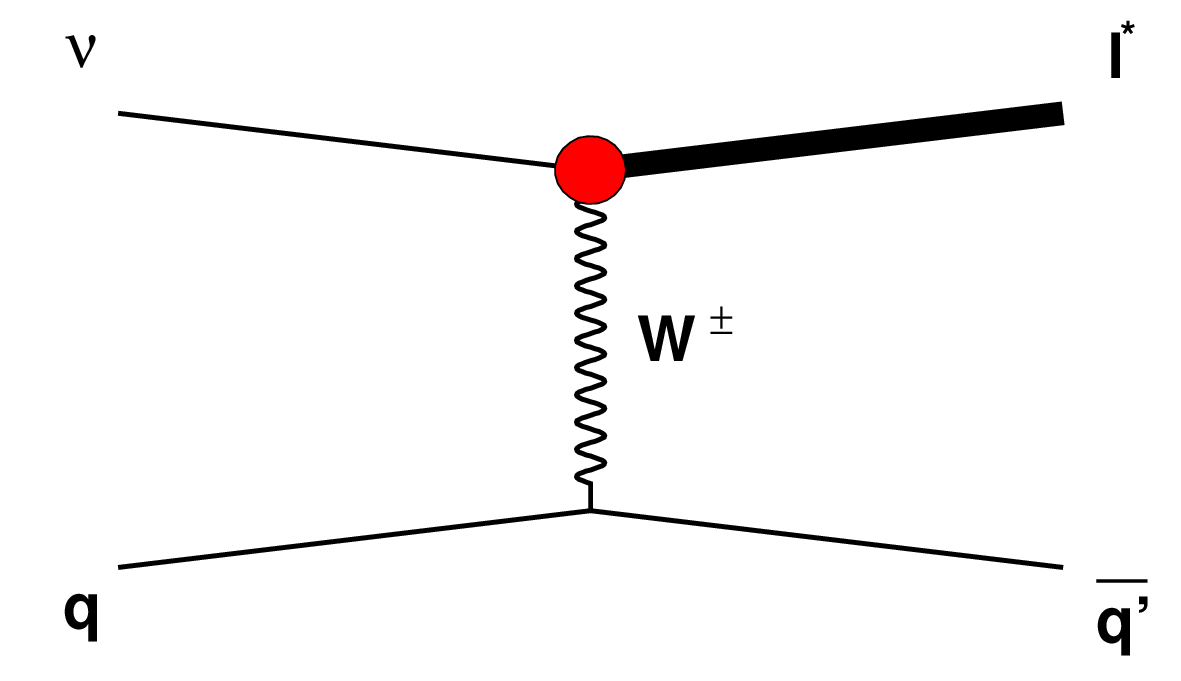,width=0.45\textwidth}}
\mbox{\epsfig{file=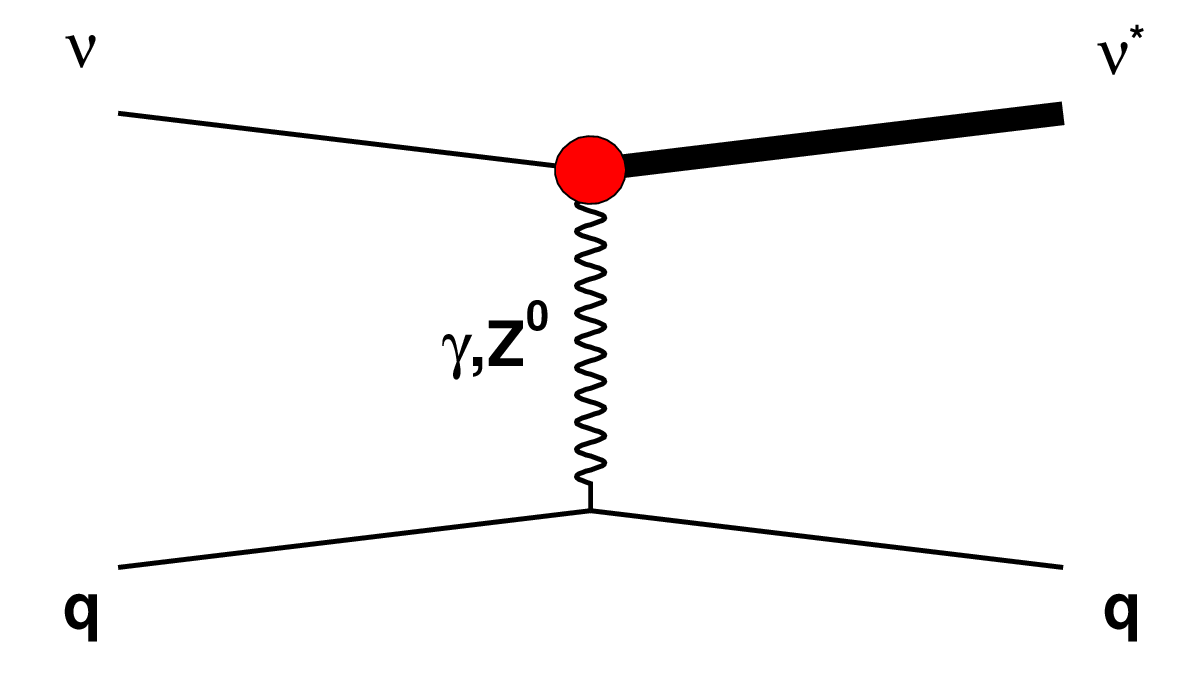,width=0.45\textwidth}}

\caption{Lowest order Feynman diagrams for the single production of excited leptons
in neutrino-quark collisions via charged current (left) and neutral current (right)
interactions. The vertex shown as a closed circle represents a $\lepton \elepton V$ coupling
(\mbox{$V \equiv \gamma,\wboson,\zboson$}) proportional to $1/{\Lambda}$.
The t-channel photon exchange can only occur in scenarios with $f \ne f'$.}
\label{fig:Feynman}
\end{center}
\end{figure}

The  production of excited leptons in neutrino-parton collisions   
is  described at the lowest order by the $t$-channel exchange of a \wboson\ boson,
in the case of excited charged lepton production (Charged Current, CC), or of a \zboson\  boson, for 
excited neutrino production (Neutral Current, NC). 
In the case of the neutral currents, 
an additional contribution from $t$-channel $\gamma$ exchange 
arises in scenarios with $f \ne f'$, due to the non-vanishing coupling to the photon. 
The tree level diagrams are shown
in Fig.\ref{fig:Feynman}. 

From the above Lagrangian, the differential cross-section for 
neutrino-parton interactions can be written as: 
\begin{equation}
\frac{d\sigma_{\nu q}}{d Q^2}(\hat s,Q^2)=2\pi\alpha^2 \left(\frac{f}{\Lambda}\right)^2 Q^2
[D_l(Q^2) S(\hat s,Q^2)\pm \bar D_l(Q^2) A(\hat s,Q^2)]
\end{equation}
\noindent
where 
the plus and minus sign apply for partons and antipartons, respectively,
$-Q^2$ is the momentum transfer, $\hat s$ is the parton level centre-of-mass
energy and:
$$
S(\hat s,Q^2)=2-(2-r)\left(\frac{Q^2}{\hat s}+r\right),~~~
A(\hat s,Q^2)=r\left(2-\frac{Q^2}{\hat s}-r\right)
$$

\noindent
where $r\equiv m_*^2/\hat s$ and $m_*^2$ is the excited lepton mass.
In the case of charged excited lepton production via CC, the $D_l$, $\bar{D_l}$ 
functions can be written as:
$$
D_e=\bar D_e=
\left(\frac{f_W}{f}\right)^2 \frac{a_W^2+v_W^2}{(Q^2+M_W^2)^2}. 
$$

For excited neutrino production via NC, both the \zboson\ and the $\gamma$
contribution have to be taken into account and the $D_l$, $\bar{D_l}$ 
can be written:
$$
D_\nu=
\frac{e_q^2}{(Q^2)^2}\left(\frac{f_\gamma}{f}\right)^2
+\frac{2 e_q v_q^Z}{Q^2(Q^2+M_Z^2)} \left(\frac{f_\gamma f_Z}{f}\right)^2
+\frac{[(v_q^Z)^2+(a_q^Z)^2]}{(Q^2+M_Z^2)^2} \left(\frac{f_Z}{f}\right)^2
$$
$$
\bar D_\nu=
\frac{2 e_q a_q^Z}{Q^2(Q^2+M_Z^2)}\left(\frac{f_\gamma f_Z}{f}\right)^2
+\frac{2v_q^Z a_q^Z}{(Q^2+M_Z^2)^2} \left(\frac{f_Z}{f}\right)^2.
$$
In the expressions above, the SM couplings are:
$$
a_W=v_W=\frac 1{2\sqrt{2} s_W}, \quad v_q^Z=\frac{2 I_{3L}^q-4 e_q s_W^2}{4 c_W s_W},
\quad a_q^Z=\frac{2I_{3L}^q}{4 c_W s_W}
$$
where $I_{3L}^q$ is the quark weak isospin.

The double differential Deep Inelastic Scattering (DIS) neutrino-nucleon cross-sections can 
be written as
\begin{eqnarray}
\frac{d^2 \sigma_{\nu N}}{dx dy}&=&\sum_{q} q(x,Q^2) \frac{d\sigma_{\nu q}}{dy}_{| \hat s=x s}
=\sum_{q} q(x,Q^2) ~xs~ \frac{d\sigma_{\nu q}}{dQ^2}_{| \hat s=x s}=\nonumber \\
&=& 2\pi\alpha^2 \left(\frac{f}{\Lambda}\right)^2\!\! (xs)^2y
\sum_q [D_l(Q^2) S(x,y)\pm \bar D_l(Q^2) A(x,y)]q(x,Q^2)
\end{eqnarray}

where $y=(E_\nu-E^*)/E_\nu$ is the inelasticity parameter 
($E_\nu$ is the incident neutrino energy and $E^{*}$ is the excited lepton energy), 
$x=Q^2/s y$ is the Bjorken variable, $q(x,Q^2)$ are the quark distribution 
functions and 
the sum runs over the quark types. We neglect the top distribution function, but we take into account
the threshold suppression of the $b\to t$ transition, using the standard
``slow-rescaling'' prescription~\cite{barnett}. 
In this work the CTEQ6-DIS parton distribution functions~\cite{cteq6} were used.

The total CC or NC production cross-sections will thus be given by:
\begin{equation}
\sigma_{\nu N}(e N\to l^* X)= \int_{(m_*^2+Q_0^2)/s}^1 dx \int_{Q_0^2/xs}^{1-r} dy \frac{d^2\sigma_{\nu N}}{dx dy}+
\sigma_{el}+\sigma_{low}
\end{equation}

\noindent
where the first term is the deep inelastic scattering (DIS) contribution, and $\sigma_{el}$ and $\sigma_{low}$
are the elastic and low inelastic contributions, respectively. These last two terms will only be important in the case
$f=-f'$, which will be discussed below.
The integration limits arise from kinematic considerations and from taking 
the parton model as valid for $Q^2>Q_0^2 \simeq 5$~GeV$^2$.

At the energies of interest, the propagator damps the cross-section for high enough $x$, effectively 
limiting its value to $x\leq M_W^2/s\simeq 10^3$~GeV$/E$(GeV).
For energies above $10^9$~GeV, we are probing values of $x$ well below the available data. 
There are several approaches for extrapolating the parton distribution functions, recently reviewed 
in~\cite{reno}. 
This leads to an uncertainty in the SM neutrino cross-section predictions of about a factor 2 for the highest energies. 
In this paper we extrapolate below $x=10^{-6}$ as described in~\cite{reno}, that is, by matching
$$
x \bar q(x,Q²)=\left(\frac{x_{min}}{x}\right)^\lambda x \bar q(x_{min},Q²).
$$
With this method we reproduce the results for the SM neutrino-nucleon
cross-section as obtained in~\cite{gandhi} within 10\%.
It should be noted however that the uncertainty in the extrapolations is much lower in 
the present case than in the SM, in light of two reasons. 
One is the kinematic constraint $x>m_*^2/s$, which, for example at $m_*=1$ TeV$/c^2$ and 
$E=10^{12}$~GeV, leads to $x$ well within the available region in CTEQ6.
The other is the $x^2$ dependence of the cross-section, 
as opposed to the SM linear dependence, damping the low $x$ contributions.

In the scenario $f=-f^{'}$, the NC photon exchange diagram is also present,
and the low $Q^2$ nucleon-parton interactions have to be taken into account. 
The differential cross-section can be written in terms of the proton structure functions
$F_1$ and $F_2$, and both the elastic and the inelastic contributions were taken into
account. In the elastic case, standard proton structure functions as described
for example in~\cite{hagiboud} were used. In the inelastic case, the
parameterisation of the structure functions described in~\cite{brasse} was taken. 

It should however be noted that the low $Q^2$ region is in the present case not as relevant as 
in~\cite{hagiboud}. In fact, provided  we are well above the kinematic limit for excited
lepton production, the DIS contribution to the cross-section is largely dominant.
However, these effects become relevant near threshold and have to be taken into account.
With the cross-section changes corresponding to the $ep\to l^* X$ case, we were able
to reproduce the results in ~\cite{hagiboud}. 

\begin{figure}[hbtp]
\begin{center}
\setlength{\unitlength}{0.0105in}%
\includegraphics[width=0.7\linewidth]{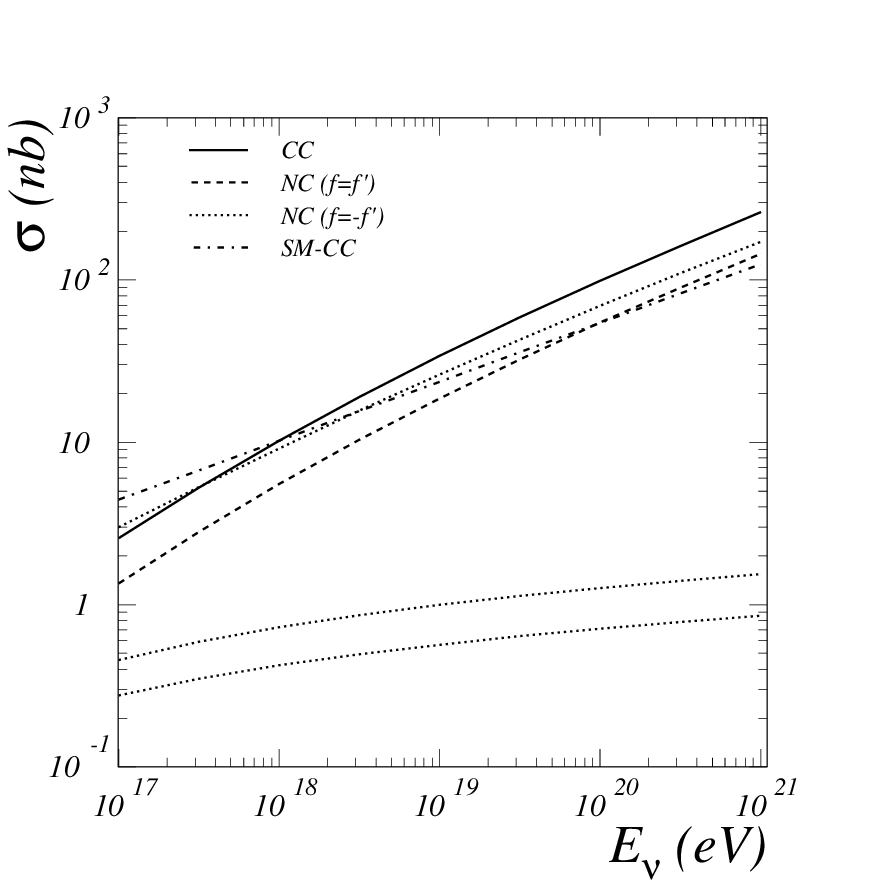}
\includegraphics[width=0.7\linewidth]{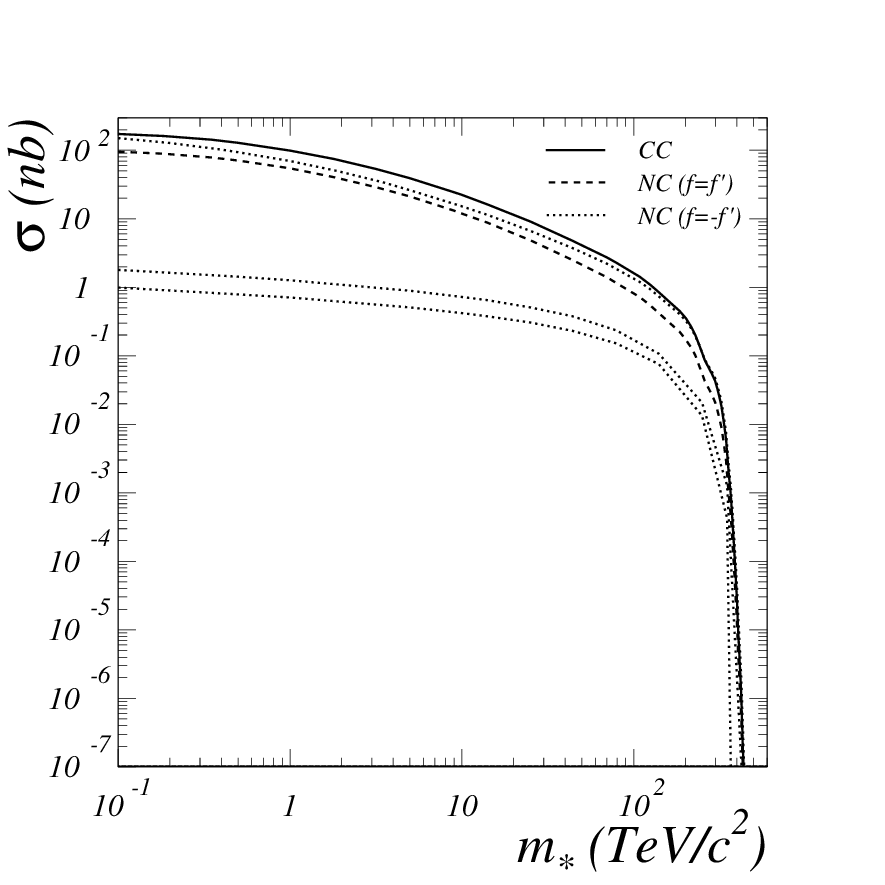}
\end{center}
\caption{Excited lepton production cross-section in $\nu N$ collisions,
via charged and neutral current interactions,  
with $f/\Lambda=15$~TeV$^{-1}$: (a) as a function of the incident
neutrino energy, for $m_*=1$~TeV$/c^2$, (b) as a
function of the excited lepton mass, for $E_\nu=10^{20}$~eV. 
The lower dotted curves show separately the elastic and inelastic low $Q^2$ 
contributions to the NC $f=-f'$ cross-section.
In (a) the SM neutrino-nucleon CC cross-section is also
shown for comparison.
}
\label{fig:sigma}
\end{figure}

The total production cross-section as a function of the incident neutrino energy is shown 
in Fig.~\ref{fig:sigma}(a), for both the charged and neutral current processes, with 
$f/\Lambda = 15$~TeV${}^{-1}$ and a chosen value of the excited lepton mass.
The total SM $\nu N$ cross-section is also shown for comparison.
In Fig.~\ref{fig:sigma}(b), the total cross-section is shown as a function of the 
excited lepton
mass, for $f/\Lambda = 15$~TeV${}^{-1}$ and a chosen value of the neutrino energy.
In both figures, the elastic and low $Q^2$ inelastic contributions to the NC, $f=-f'$
cross-section are shown separately.

\begin{figure}[hbtp]
\begin{center}
\setlength{\unitlength}{0.0105in}%
\includegraphics[width=0.7\linewidth]{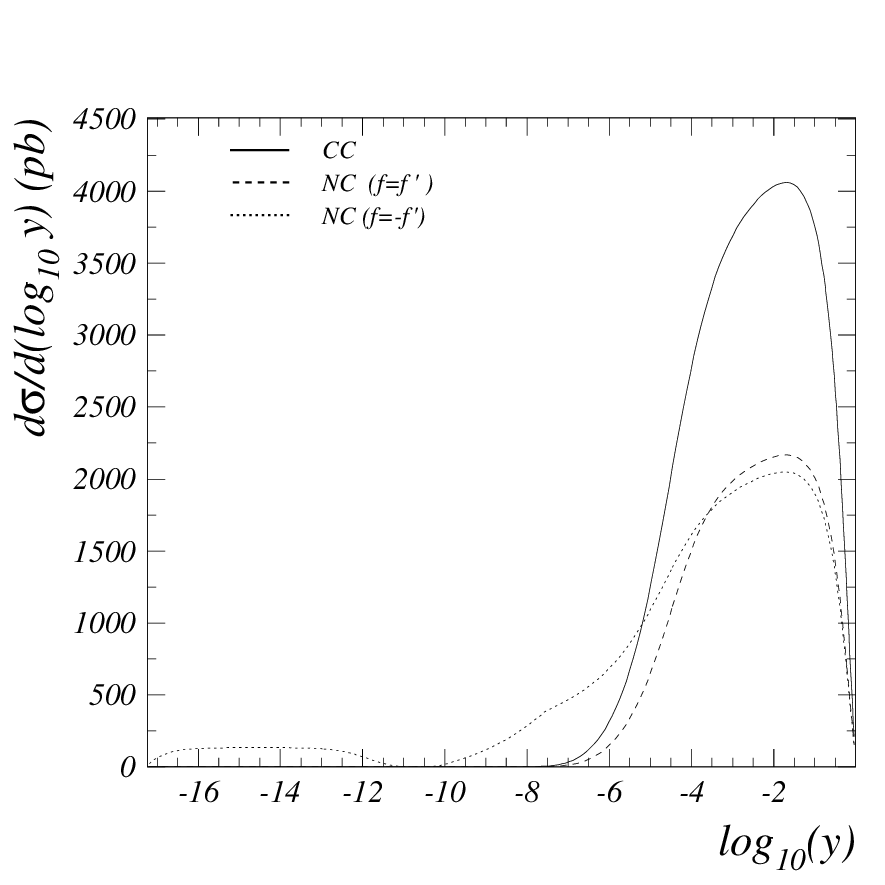}
\end{center}
\caption{Differencial excited lepton production cross-section in 
charged and neutral current $\nu N$ interactions 
with $f/\Lambda=15$~TeV$^{-1}$, 
$E_\nu=10^{20}$~eV and $m_*=14$~TeV$/c^2$.
For NC and $f=-f'$, the elastic and low
$Q^2$ inelastic contributions are visible, in the lowest and intermediate range of 
$log_{10}(y)$, respectively.}
\label{fig:dsigdy}
\end{figure}

In Fig.~\ref{fig:dsigdy}, the differential cross-sections:
\begin{equation}
\label{eq:dsigdy}
\frac{d\sigma_{\nu N}}{dy}=\int dx \frac{d^2 \sigma_{\nu N}}{dx dy}
\end{equation}
are shown as an example for fixed values of the 
incident neutrino energy and of the excited lepton mass and coupling parameters.
The charged current cross-section is shown, together with the neutral current
cross-sections for $f=f'$ and  $f=-f'$. In the latter, the elastic and low
$Q^2$ inelastic contributions are visible, in the lowest and intermediate range of 
$log_{10}(y)$, respectively.

These distributions determine the fraction of the incident neutrino energy carried
away by the hadronic component $X$ and thus, to some extent, the energy of
the observable extensive air shower. The observability of the excited lepton
decay products will depend on the decay mode, as discussed in detail
in section~\ref{sec:sensitivities}.

\subsection{Decay}
\label{sec:dec}
 
Excited leptons are assumed to decay promptly 
by radiating a $\gamma$, \wboson\   or \zboson\ boson. 
For $\Lambda = 1$~TeV and $E<10^{21}$~eV, their decay length is predicted to be less than $10^{-4}$~m 
and, in all the studied scenarios, they decay essentially at the production point.
The decay branching ratios are also functions of the $f$ and $f'$ parameters. 
While at lower masses the branching ratios show an important dependence on the 
excited lepton mass, they are practically constant in the interesting mass
range ($m_*>200$~GeV$/c^2$).
For charged excited leptons, the electromagnetic radiative decay is forbidden
if \mbox{$f=-f'$} and the decays proceed exclusively 
through  \zboson\ and \wboson\ bosons, with branching fractions of about 40\% and
60\%, respectively.  
However, as long as \mbox{$f \neq - f'$}, there is a significant contribution to the 
total decay width from the electromagnetic radiative decay, even if the difference  
$|f|-|f'|$ is much smaller than $|f|$. 
In the \mbox{$f=f'$} case, the electromagnetic radiative decay is largely 
dominant at masses below the \wboson, \zboson\ gauge boson masses. 
In the presently interesting mass range,
the decay into the \wboson\ is again about 60\%, while the branching
ratios of the decays through a photon or a \zboson\ are of the order of 30\%
and 10\%, respectively. 

In cosmic ray air shower experiments, only the excited lepton decay products 
originating hadronic or electromagnetic showers will contribute to the 
EAS. Thus, hadronic jets from the decay of heavy gauge bosons, electrons,
and photons will contribute to the shower energy, while final state neutrinos
and muons will go undetected. 

High energy taus may produce double bang signatures of the type described 
in~\cite{dbang,bh-our}.
In fact, in the relevant energy range, taus have an interaction length in air which
is much larger than their decay length, and a decay length large enough for the
production of a well separated second bang - a second shower produced by its decay.

\section{Limits and sensitivities}
\label{sec:sensitivities}

The expected number of observed excited lepton events is given by:
\begin{equation}
\label{eq:nevents}
{\mathscr N} = N_A \int~\frac{d\phi_{\nu}}{dE_{\nu}}~\sigma_{{\nu N}}~{\mathscr A}~\Delta T~dE_{\nu},
\end{equation}
where $d\phi_{\nu}/dE_{\nu}$ is the incident neutrino flux,
$\sigma_{{\nu N}}$ is the neutral or charged current production
cross-section, depending on whether
neutral or charged excited lepton production is considered, 
${\mathscr A}$ is the acceptance 
of the experiment for the extensive air showers produced by these final states,
$\Delta T$ is the observation time interval and $N_A$ is Avogrado's number.
It is assumed that the attenuation of neutrinos in the atmosphere can be neglected,
which is a safe assumption for total neutrino-nucleon cross-sections 
in the range relevant for the present study. For larger values of
the cross-section, the treatment discussed in~\cite{agasa-thesis}
should be applied.
The estimation of the different factors in the expression is discussed below.

In this work the Waxman-Bahcall (WB)~\cite{wb} bound with no z evolution, 
$E_\nu^2 \frac{d\phi}{dE_\nu} = 10^{-8}$ [GeV/cm$^2$ s sr], is assumed. 
This flux is much lower than the 
cascade~\cite{cascade} or the Mannheim-Protheroe-Rachen upper bounds~\cite{mpr} and is, 
in the relevant neutrino energy range, 
higher but of the same order of magnitude of the ``best prediction'' computation
for cosmogenic neutrinos presented in~\cite{agasa-thesis}.
Taking into account the existence of neutrino oscillations over cosmological distances,
equal flux (one third of the WB flux) for each neutrino flavour was considered.

The neutral and charged current excited lepton production cross-sections
in neutrino-nucleon collision processes $\nu N \rightarrow \nu^* X$ and
$\nu N \rightarrow \ell^* X$ are the ones computed in section~\ref{sec:prod}.

The observation times were assumed to be: 10 years for Auger, 3 years and 10\% duty 
cycle for both EUSO and OWL. For Agasa and Fly's Eye, we followed 
reference~\cite{agasa-a}.

\subsection{Acceptance}

\begin{figure}[hbtp]
\begin{center}
\setlength{\unitlength}{0.0105in}%
\includegraphics[width=0.6\linewidth]{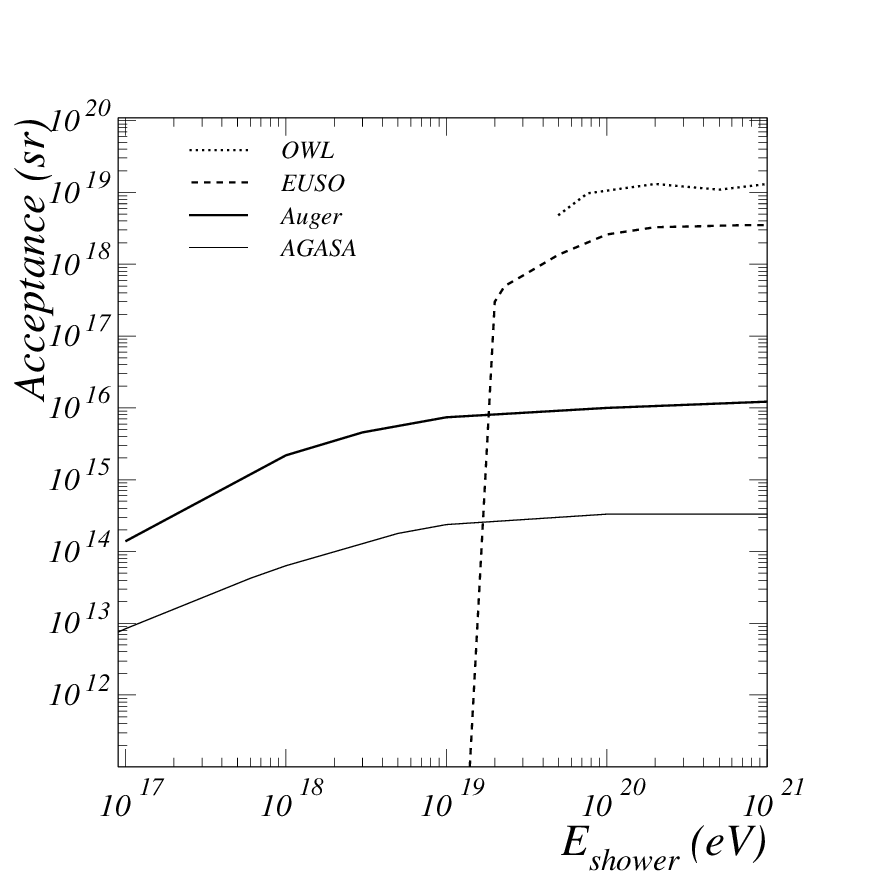}
\includegraphics[width=0.6\linewidth]{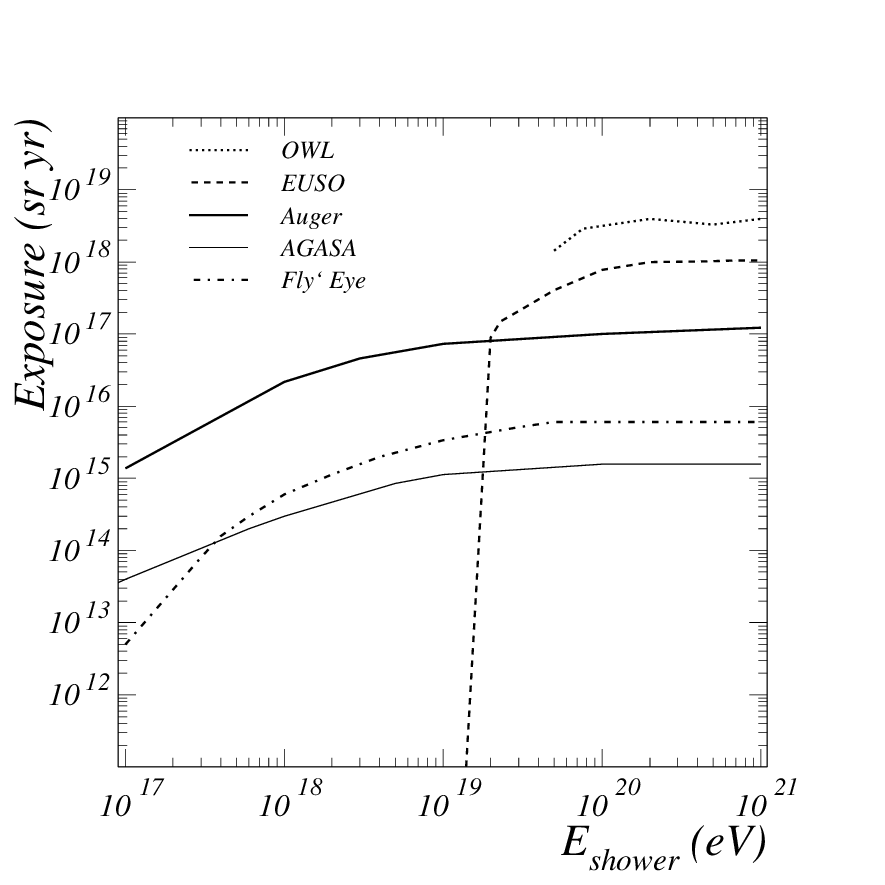}
\end{center}
\caption{(a)Acceptances and (b) Exposures of air shower cosmic ray
experiments as a function of the final state extensive air
shower energy. The information in references~\cite{auger-a,agasa-a,euso-a,owl-a} 
was used. 
The observation times were taken as: 10 years for Auger, 3 years and 10\% duty cycle
for both EUSO and OWL. For Agasa and Fly's Eye, we followed reference~\cite{agasa-a}.
It should be noted that the relation between the 
shower energy and the incident neutrino energy is process
dependent.}
\label{fig:accept}
\end{figure}

The acceptance ${\mathscr A(E)}$ in equation~\ref{eq:nevents} includes both the geometrical 
aperture, the target density and the detection efficiency factors:
$$
{\mathscr A(E)} = \int~\rho(\ell)~A(E) cos \theta~\epsilon(E)~\Delta \Omega~d \ell, 
$$
where $A(E)cos \theta$ is the effective area, $\rho(\ell)$ is the atmospheric density profile, 
$\epsilon(E)$ is a global detection efficiency factor and 
$\Delta \Omega$ is the observation solid angle.
Under similar assumptions,
acceptances have been computed for different experiments 
in the context of the estimation of the sensitivity for cosmic 
neutrinos~\cite{agasa-a,auger-a,euso-a,owl-a}. 
It should be noted that, whereas these acceptances are valid for any $\nu N$ 
interaction process, the relation between the shower energy and the primary neutrino
energy is process dependent. 
Taking the SM as an example, 
whereas in the charged current process $\nu_e N \rightarrow e N$ the 
energy of the observed extensive air shower corresponds to the energy of the incident
neutrino, this is not the case for the remaining neutrino families and for
the neutral current case, $\nu N \rightarrow \nu N$, in which the final state neutrino
goes undetected. The fraction of the primary energy that goes into 
EAS energy in the SM NC process is of the order of 20\%
for energies around $10^{19}$~eV, and decreases slowly with energy. 
It is thus convenient for the present purposes to plot the neutrino acceptances of 
the different experiments as a function of the shower energy.
These acceptances are compiled in Fig.~\ref{fig:accept}(a).
For Auger, the most conservative estimate of the acceptance for quasi-horizontal
showers as determined by Monte Carlo simulations~\cite{auger-a} was considered.
The acceptance of AGASA was conservatively taken as the acceptance 
for electromagnetic showers given in~\cite{agasa-a}.
A discussion of the acceptance of AGASA for hadronic and electromagnetic
showers can be found in~\cite{agasa-thesis}.
The acceptance of EUSO was taken from the estimation in~\cite{euso-a}, including
the trigger and visibility of the shower maximum conditions
but no cloud effect (which was included as a reduction of the duty cycle).
The acceptance of OWL was estimated from the aperture given in~\cite{owl} for an
altitude of 500 Km. 

In Fig.~\ref{fig:accept}(b) the exposures of the different experiments are shown,
which take into account both the acceptance and the effective observation time.
The assumed observation periods are the ones described above (and quoted
in the caption of the figure).

In the case of excited lepton production, $\nu N \rightarrow \nu^* X$ or 
$\nu N \rightarrow \ell^* X$, 
the relation between the shower energy and the incident neutrino 
energy is obtained from the $d\sigma_{\nu N}/dy$ distribution (see section~\ref{sec:excited},
equation~\ref{eq:dsigdy}) and depends on the 
decay mode of the produced neutral or charged excited lepton.
The fraction of the incident neutrino energy carried
away by the hadronic component $X$ and thus, to some extent, the energy of
the observable extensive air shower, are determined by these
distributions (such as the one in Fig.~\ref{fig:dsigdy}).
On the other hand, the observability of the excited lepton
decay products will depend on the decay mode.
For this reason, average values of the acceptance were computed via Monte Carlo,
in the way detailed below.

\begin{figure}[hbtp]
\begin{center}
\setlength{\unitlength}{0.0105in}%
\includegraphics[width=0.7\linewidth]{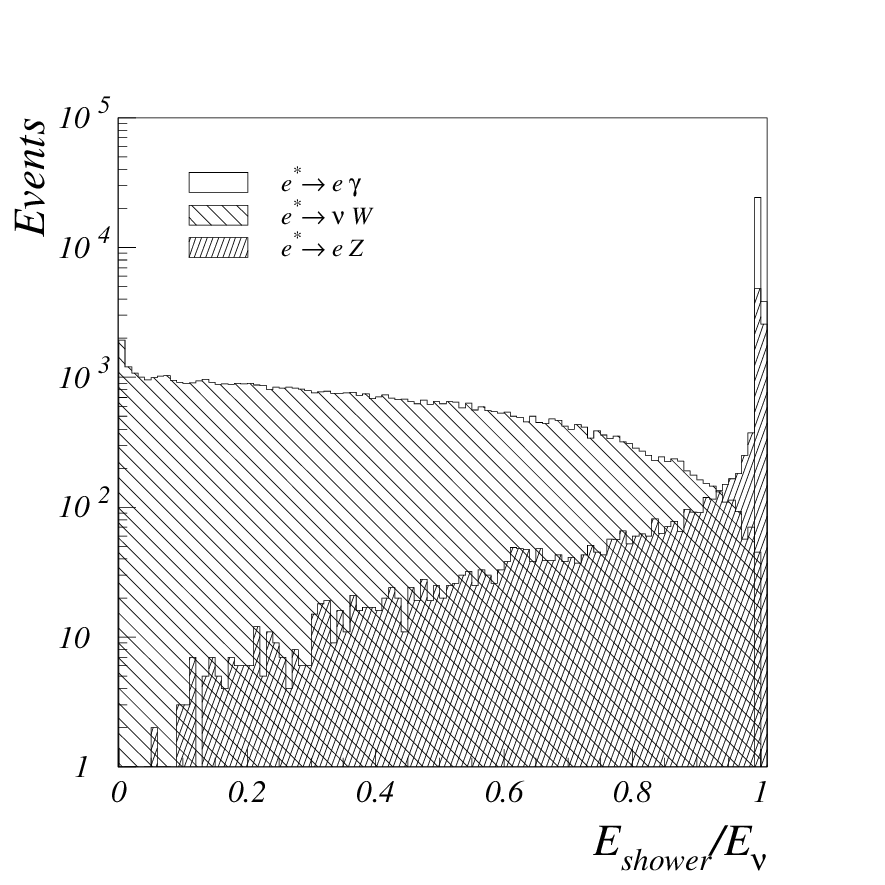}
\includegraphics[width=0.7\linewidth]{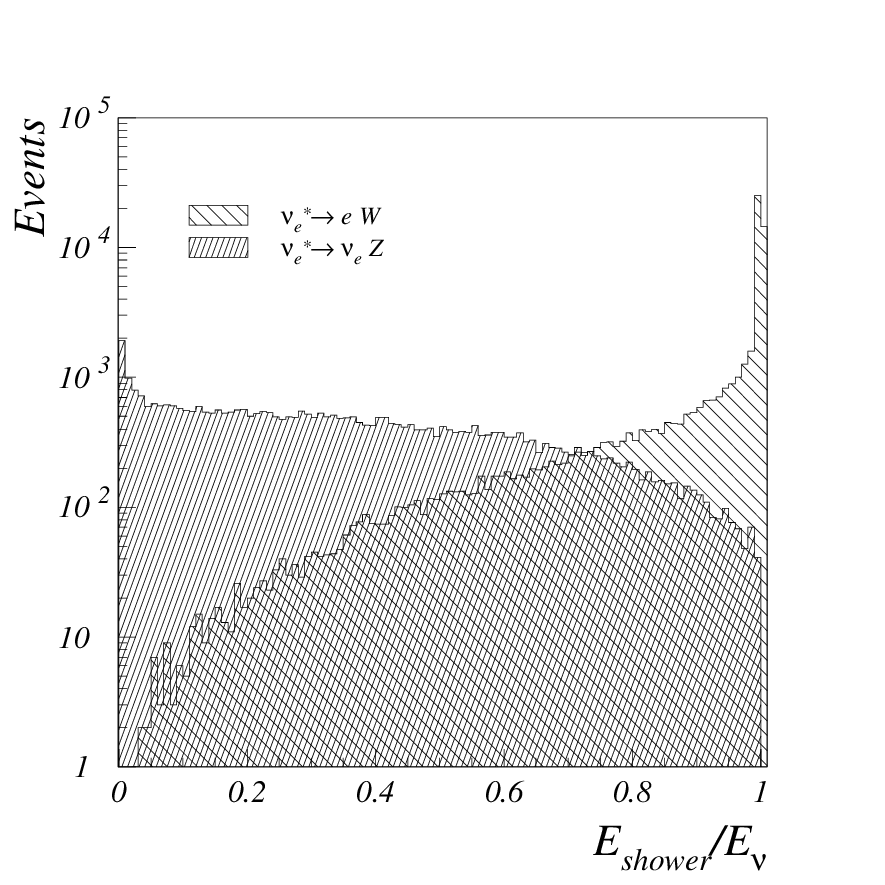}
\end{center}
\caption{Ratio of the shower energy to the incident neutrino energy
in the different decay modes of the excited lepton
for (a) excited electron production and (b) excited electron neutrino
production, with $f=f'$, $m_*=1$~TeV$/c^2$ and $E_\nu=10^{20}~eV$.  
}
\label{fig:edist}
\end{figure}

\begin{figure}[hbtp]
\begin{center}
\setlength{\unitlength}{0.0105in}%
\includegraphics[width=0.7\linewidth]{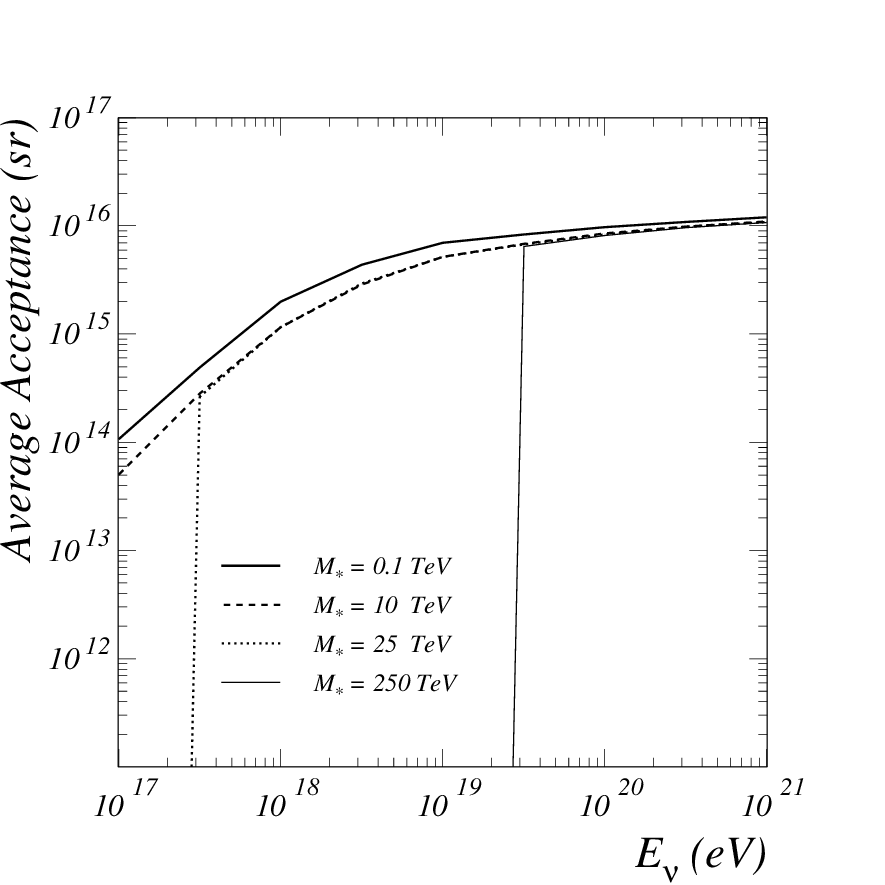}
\includegraphics[width=0.7\linewidth]{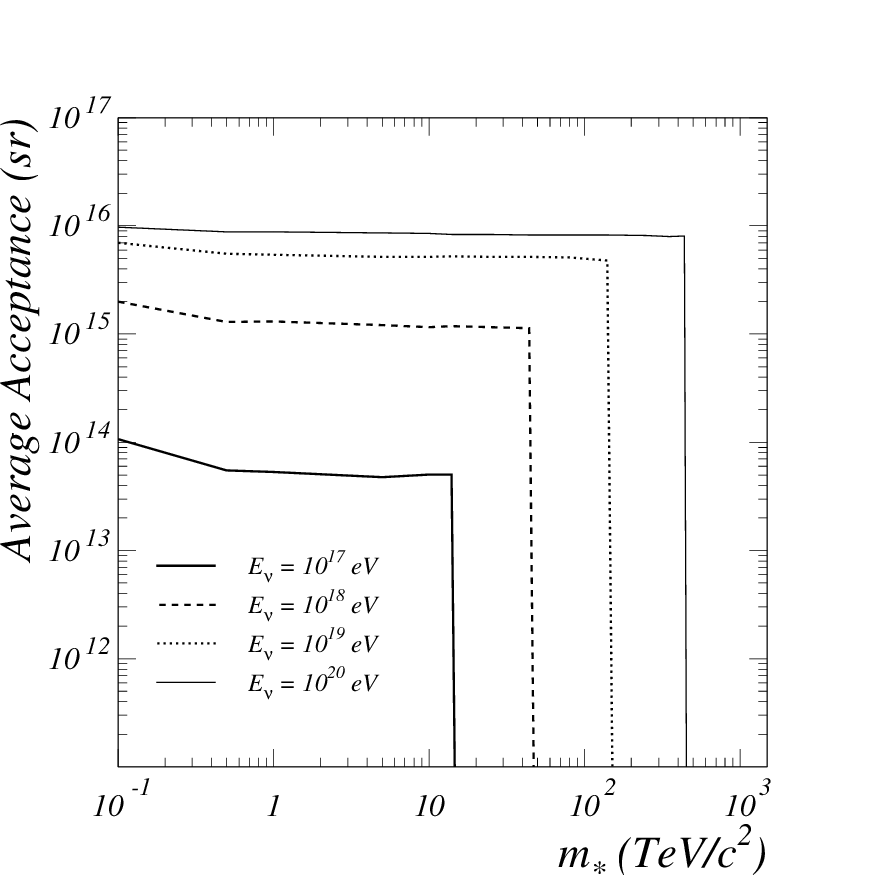}
\end{center}
\caption{Average acceptance computed for Auger: (a) as a function
of the neutrino energy, for different values of the excited lepton
mass; (b) as a function of the excited lepton mass, for different
values of the incident neutrino energy. In this example, excited electron 
production with $f=f'$ was considered.
}
\label{fig:avaccept-auger}
\end{figure}

\begin{figure}[hbtp]
\begin{center}
\setlength{\unitlength}{0.0105in}%
\includegraphics[width=0.7\linewidth]{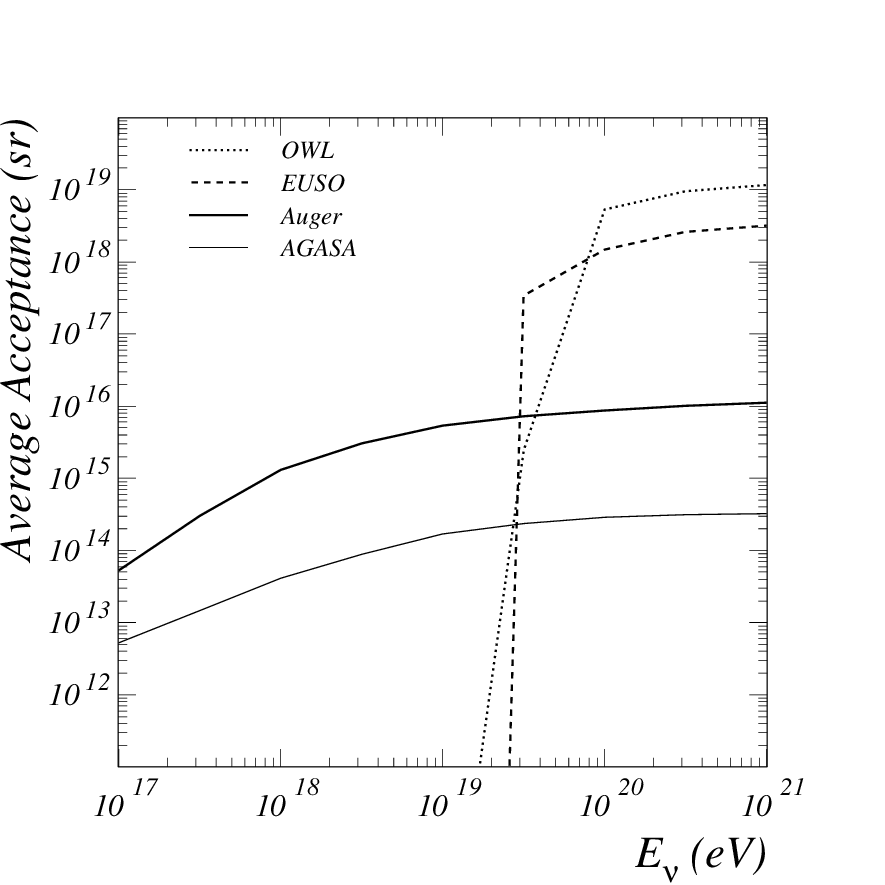}
\includegraphics[width=0.7\linewidth]{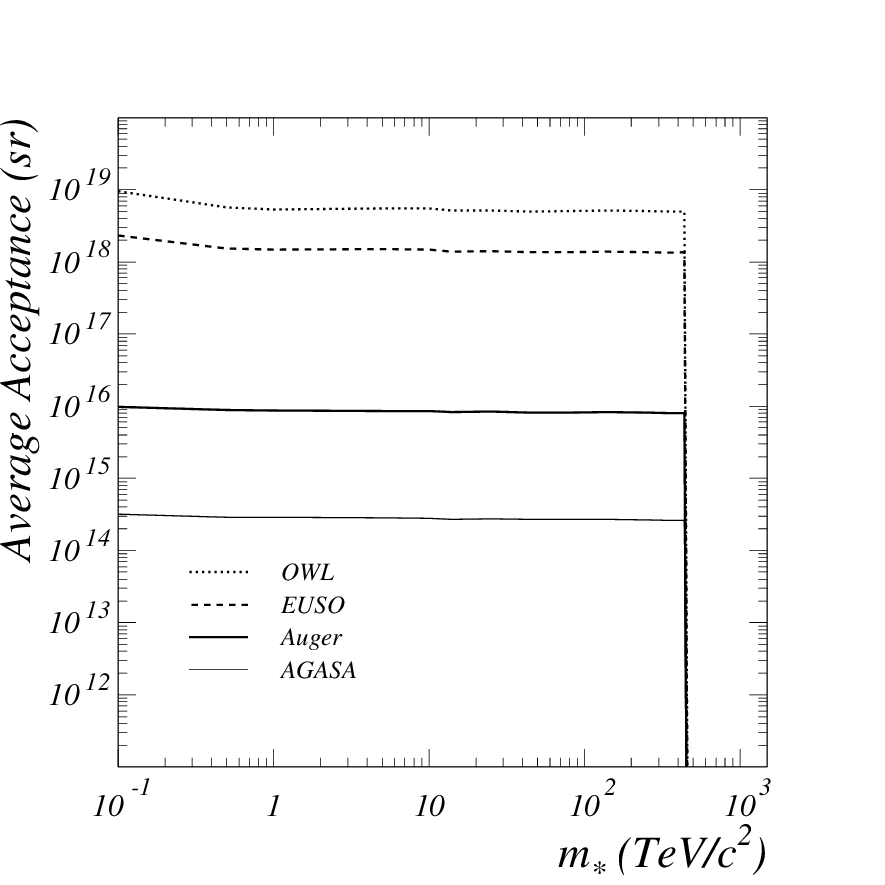}
\end{center}
\caption{Average acceptances of the different experiments
as a function of (a) the incident neutrino energy, for
$m_*=1$~TeV$/c^2$; (b) the excited lepton mass, with $E_\nu=10^{20}$~eV.
These results are for
excited electron production with $f=f'$.
}
\label{fig:avaccept-all}
\end{figure}

For given values of the incident neutrino energy, the excited lepton mass and 
the $f$, $f^{'}$ parameters, the production and decay of the excited leptons were
simulated, taking into account the  $d\sigma_{\nu N}/dy$ distributions and the branching
ratios of the excited leptons. 
In this generator, the model described in section~\ref{sec:excited} 
for excited lepton production and decay was implemented.
The decay of the heavy gauge bosons arising from
the excited lepton decay, as well as the hadronisation of the final state
quarks, were handled by JETSET~\cite{jetset}.
For each set of input parameter values, one 
thousand events were generated for each excited lepton type.
For each case, the shower energy
was computed (as described in section~\ref{sec:dec}). 
The ratio of the shower energy to the incident neutrino energy
in the different decay modes of the excited leptons is shown for example
cases in Fig.~\ref{fig:edist}.
The corresponding acceptances were then obtained from Fig.~\ref{fig:accept}.
Averaging over all the generated events,
an average acceptance was determined for each set of 
input parameters, as a function of the incident neutrino energy.
The average acceptance computed for Auger is shown in Fig.~\ref{fig:avaccept-auger}
as a function of the incident neutrino energy and of the excited lepton mass.
It can be seen that while the kinematic limit effect is clearly visible,
at higher energies the average acceptances follow relatively closely 
the one shown in Fig.~\ref{fig:accept}. 
The average acceptances obtained for the different experiments 
are shown in Fig.~\ref{fig:avaccept-all}, for $E_\nu=10^{20}~eV$ and $m_*=1$~TeV$/c^2$.
Again, the average acceptances are close to 
the ones shown in Fig.~\ref{fig:accept}. 
This is due to the fact that in the dominant decay modes the fraction of the
neutrino energy visible as shower energy is relatively high.

\subsection{Results}

Using equation~\ref{eq:nevents}, the sensitivity of the different experiments 
to excited lepton production, as a function of the excited lepton mass, was
studied. Requiring the observation of one event, the sensitivity on the 
ratio $f/\Lambda$ (see section~\ref{sec:excited}) as a function of the mass was derived.

\begin{figure}[hbtp]
\begin{center}
\setlength{\unitlength}{0.0105in}%
\includegraphics[width=0.495\linewidth]{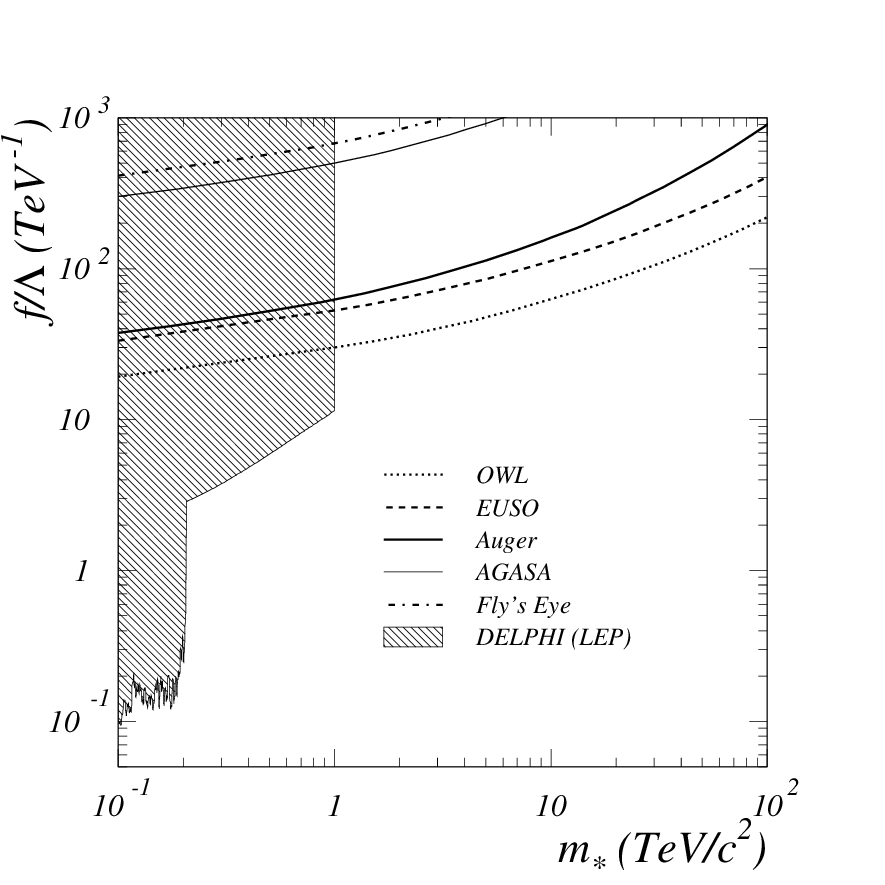}
\includegraphics[width=0.495\linewidth]{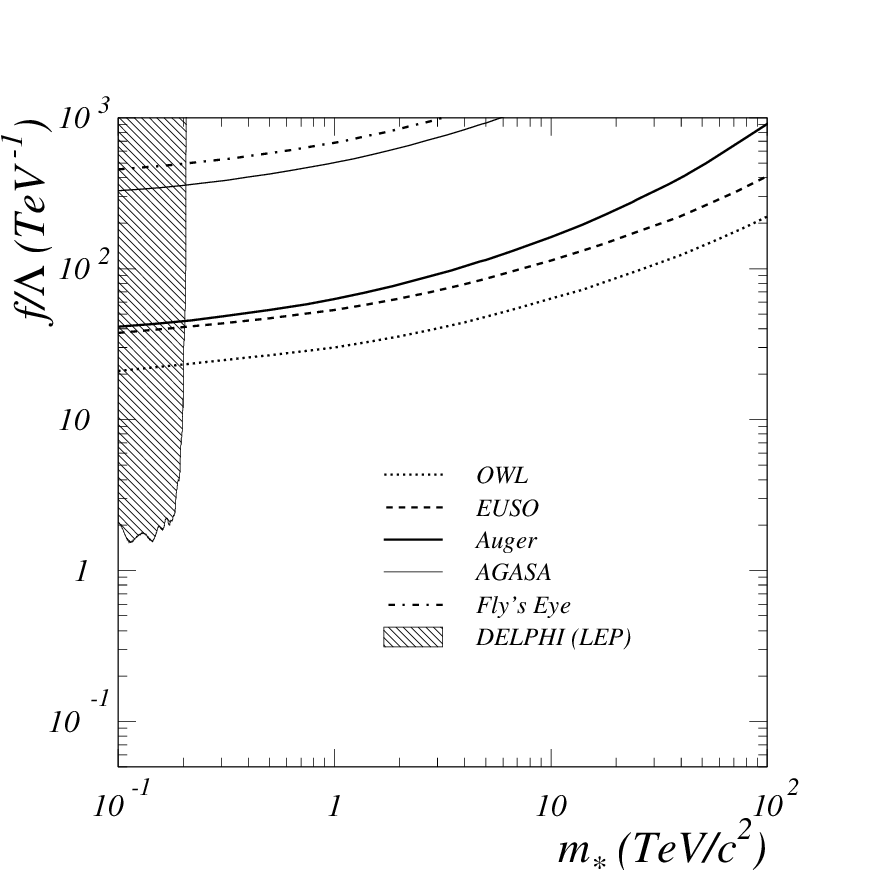}
\includegraphics[width=0.495\linewidth]{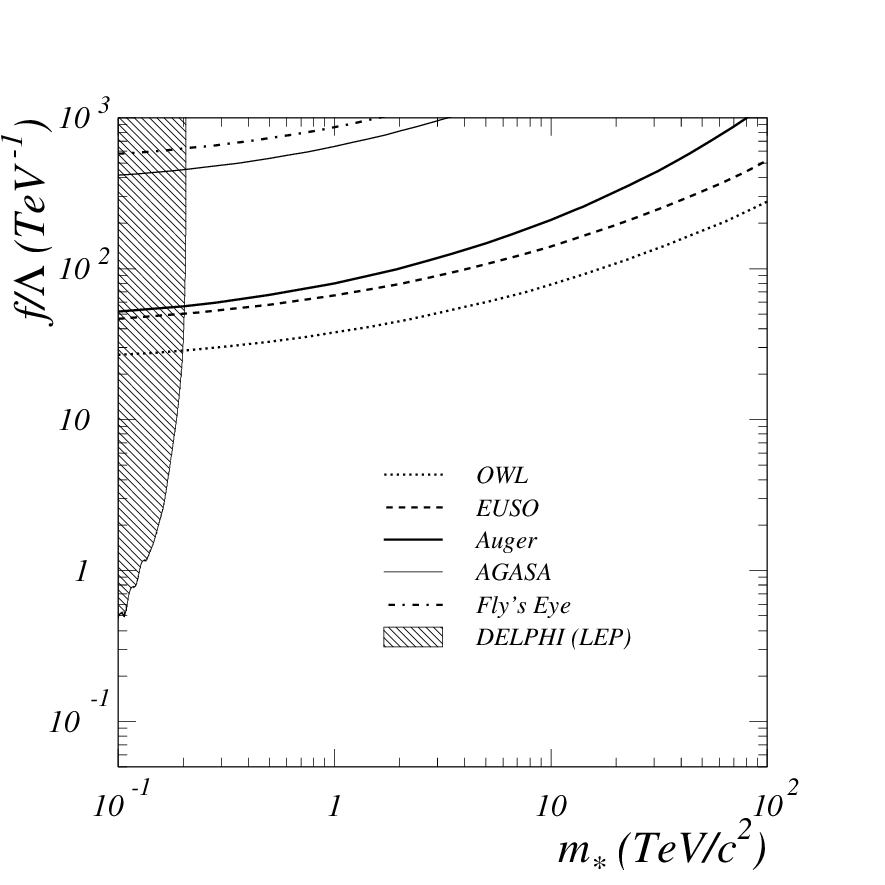}
\includegraphics[width=0.495\linewidth]{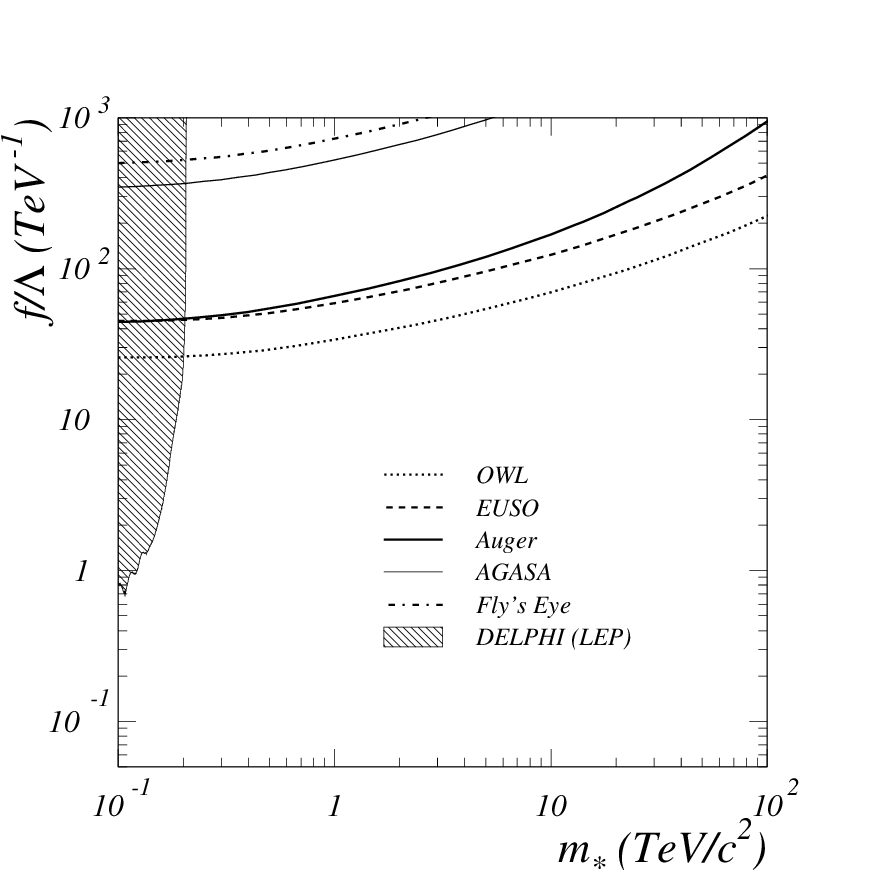}
\end{center}
\caption{Estimated sensitivities of the different
experiments as a function of the excited lepton
mass, for excited electrons (upper plots) and 
excited electron neutrinos (lower plots), 
in the scenarios $f=f'$ (left) and $f=-f'$ (right).
The regions excluded by LEP are also shown (in dashed) for comparison.
The observation times were taken as: 10 years for Auger, 3 years and 10\% duty cycle
for both EUSO and OWL. For Agasa and Fly's Eye, we followed reference~\cite{agasa-a}.
}
\label{fig:results1}
\end{figure}

Fig.~\ref{fig:results1} shows the obtained 
sensitivities for first family excited leptons (excited electrons and excited electron neutrinos) 
as a function of the excited lepton
mass, for the scenarios $f=f'$ and $f=-f'$.
The assumed observation times are the ones detailed above (and quoted
in the caption of the figure).
For comparison, limits on $f/\Lambda$ obtained in direct and indirect searches 
for excited leptons at LEP are also shown~\cite{lep}.
LEP direct searches exclude $f/\Lambda$ values down to below 1 TeV$^{-1}$,
in a mass range that extends up to about 200 GeV. 
Indirect searches, only applicable for excited electrons with non-vanishing
electromagnetic coupling ($f=f'$ in our case) extend the exclusion to much
higher masses, although with a poorer sensitivity.

These results show that, for the foreseen acceptances,
observation time intervals and fluxes, cosmic ray experiments may detect excited
lepton production only for rather large values of the coupling $f/\Lambda$.
In this situation, they will nevertheless greatly extend the 
mass region explored at accelerators.

\begin{figure}[hbtp]
\begin{center}
\setlength{\unitlength}{0.0105in}%
\includegraphics[width=0.495\linewidth]{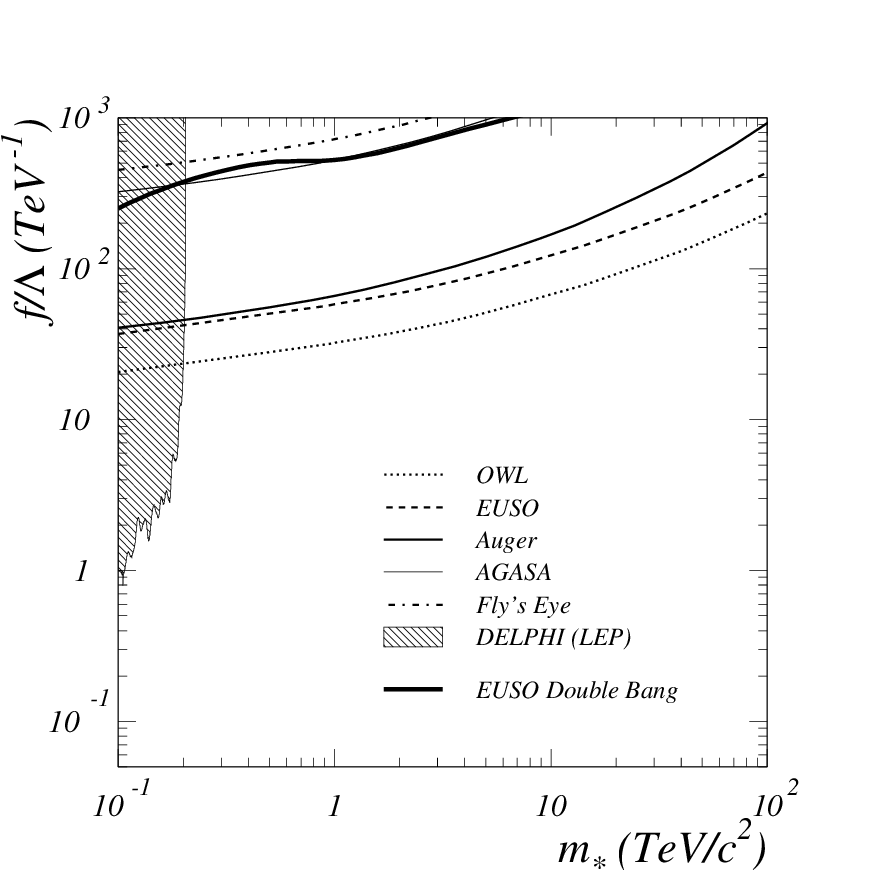}
\includegraphics[width=0.495\linewidth]{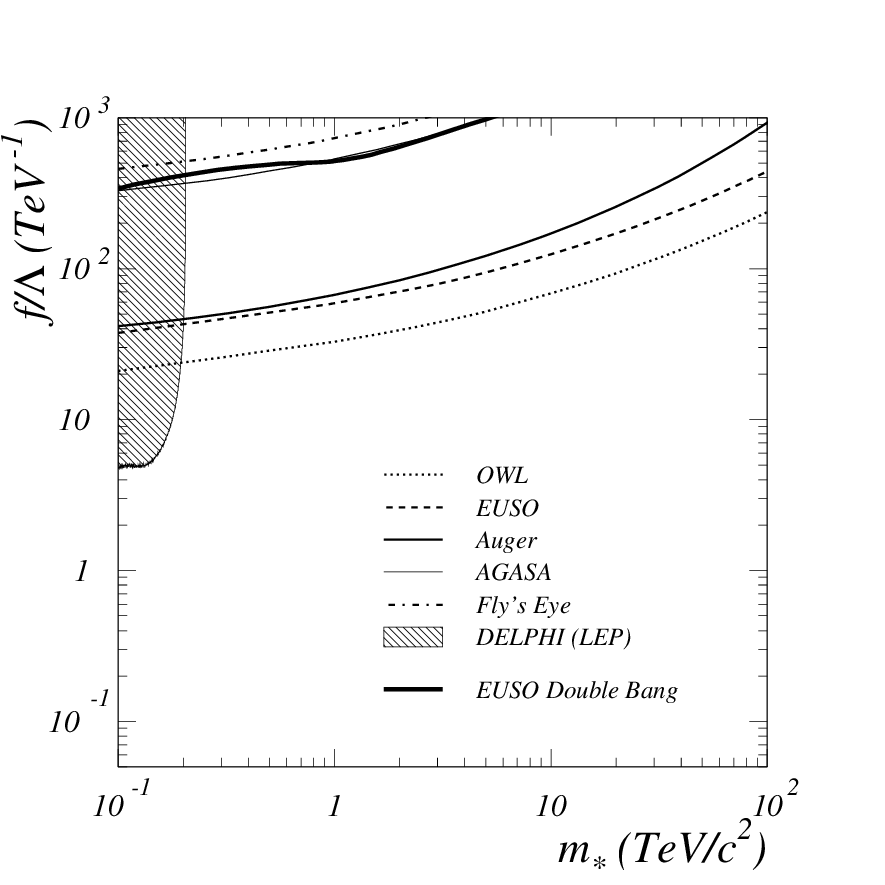}
\includegraphics[width=0.495\linewidth]{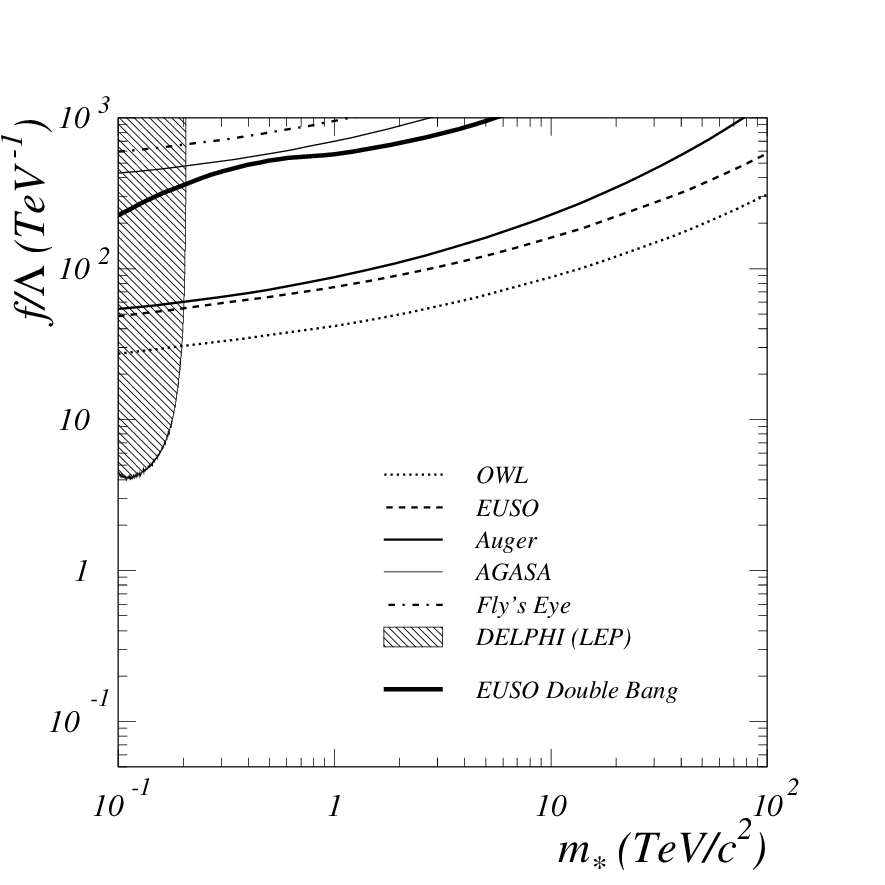}
\includegraphics[width=0.495\linewidth]{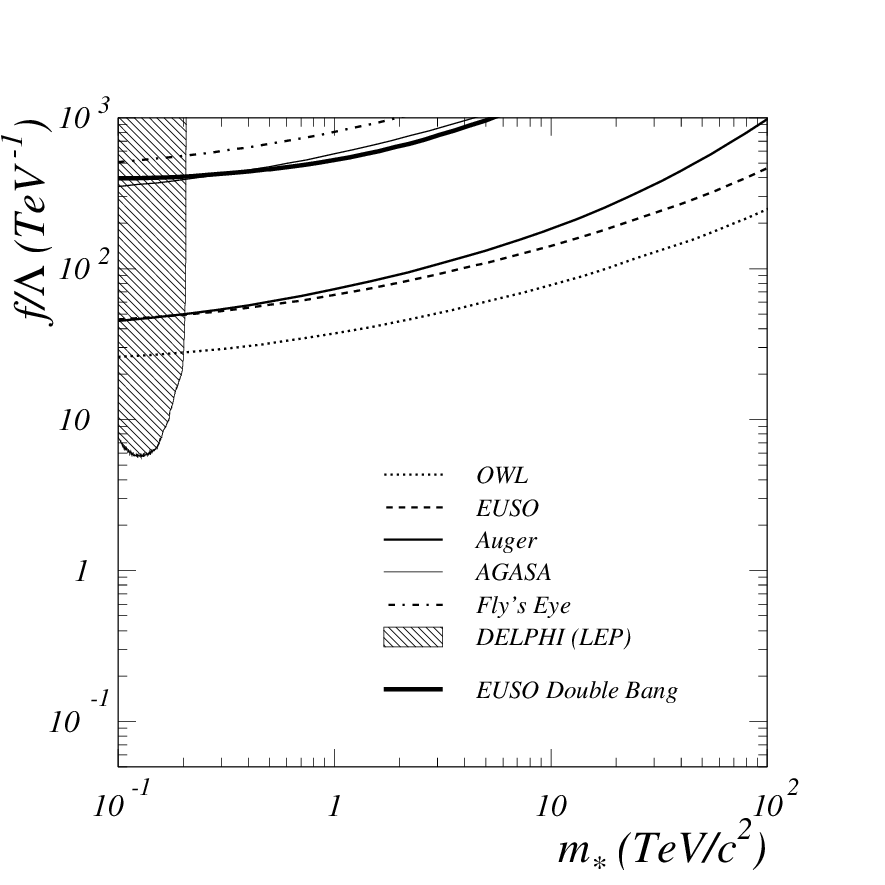}
\end{center}
\caption{Estimated sensitivities of the different
experiments as a function of the excited lepton
mass, for excited taus (upper plots) and 
excited tau neutrinos (lower plots), 
in the scenarios $f=f'$ (left) and $f=-f'$ (right).
The regions excluded by LEP are also shown (in dashed) for comparison.
The observation times were taken as: 10 years for Auger, 3 years and 10\% duty cycle
for both EUSO and OWL. For Agasa and Fly's Eye, we followed reference~\cite{agasa-a}.
The sensitivity curves from  double bang events in EUSO are also shown. 
}
\label{fig:results3}
\end{figure}

Excited leptons of different flavours, both charged and neutral,
were also studied.
The obtained sensitivities are, for these cases, comparable but
slightly worse, due to the lower shower energy, for the same energy
of the incident neutrino. 
The results obtained for third family excited leptons are shown
in Fig.~\ref{fig:results3}.

Furthermore, for third family excited leptons, both charged and
neutral, an energetic
tau lepton could be produced in the decay.
In this case, the double bang signature proposed in~\cite{dbang,bh-our}
could be searched for: the tau could travel long enough for its
decay to produce a second shower, separate from the first one, 
but still within the field of view of the experiment. 
This rather distinctive new physics signature obviously requires
a very large field of view, while the energy threshold for the
observation of the second bang is also a critical issue. 
In fact, even for the experiments with the largest acceptances,
only a few percent of the detected events are expected to 
have a visible second bang. Using the procedure detailed 
in~\cite{bh-our}, the sensitivity from the observation
of double bang events in EUSO was estimated. This curve 
is also shown in Fig.~\ref{fig:results3}, where we see that
we loose about one order of magnitude with respect to the
sensitivity of the experiment. 

\section{Conclusions}
\label{sec:concl}

Excited leptons could be produced in the interaction of high energy quasi-horizontal 
cosmic neutrinos with the atmosphere via neutral and charged current processes,
$\nu N \rightarrow \nu^* X$ and $\nu N \rightarrow \ell^* X$.
The possibility of detecting excited lepton in present and future very high 
energy cosmic ray experiments was studied.

The model in~\cite{hagiboud} was used to compute
charged and neutral current cross-sections for the very high energy
range. The results were cross-checked with those given in~\cite{hagiboud} for
$ep$ collisions at much lower energies, and also with calculations of the
SM $\nu N$ cross-section at very high energies.

Monte Carlo methods were used to estimate the average acceptances, as a function 
of the neutrino energy, taking into account
the computed differential cross-section and the excited lepton decay
branching-ratios, as well as the neutrino acceptances of present and future very 
high energy cosmic ray experiments quoted in the literature~\cite{agasa-a,auger-a}.

Sensitivity curves in the coupling $f/\Lambda$ as a function of the excited 
lepton mass were obtained for different experiments,
assuming the Waxman-Bahcall bound for the neutrino flux, considering charged
and neutral excited leptons of first, second and third family. 
The results show that cosmic ray air shower experiments may represent a window for
excited lepton searches in a mass range well beyond the TeV, if the coupling
$f/\Lambda$ is of the order of some tens of TeV$^{-1}$.

However, an excited lepton signal would only correspond to an increase on the
number of expected neutrino-induced horizontal showers. Although this is 
in itself a relevant prediction, the large uncertainties on the fluxes 
stress the need for complementary signatures, such as the double bang signature 
discussed above.



\newpage

\end{document}